\documentclass[12pt,letterpaper]{article}

\usepackage[margin=1in]{geometry}
\usepackage{times}
\usepackage{setspace}
\usepackage{amsmath,amssymb}
\usepackage{booktabs}
\usepackage{siunitx}
\usepackage{graphicx}
\usepackage{caption}
\usepackage{url}
\usepackage{enumitem}
\usepackage{microtype}
\usepackage{multicol}

\usepackage{natbib}
\setcitestyle{authoryear,round,semicolon,aysep={}}

\usepackage[hidelinks]{hyperref}
\hypersetup{pdftitle={Toward a Time-Aware Assessment Framework for the Carbon Cost of AI-Enabled Decarbonization},pdfauthor={Chenrui Xu; Burcu Akinci; Christopher McComb},pdfsubject={Author manuscript aligned with the final i3CE 2026 Word manuscript}}

\usepackage{fancyhdr}
\usepackage{titlesec}
\titlespacing*{\section}{0pt}{10pt}{0.4ex}
\titlespacing*{\subsubsection}{0pt}{8pt}{4pt}
\setlist[enumerate]{itemsep=4pt,topsep=4pt,parsep=0pt}

\usepackage{titling}
\pretitle{\bfseries\fontsize{12}{12}\selectfont}
\posttitle{\par}

\titleformat{\section}{\normalfont\bfseries\fontsize{12}{14}\selectfont}{\thesection}{1em}{}
\titleformat{name=\section,numberless}{\normalfont\bfseries\fontsize{12}{12}\selectfont}{}{0pt}{}

\begin{document}

\begin{center}
{\bfseries Toward a Time-Aware Assessment Framework for the Carbon Cost of AI-Enabled Decarbonization}\par
\vspace{4pt}
Chenrui Xu$^{1}$; Burcu Akinci$^{2}$; and Christopher McComb$^{3}$\par
\vspace{6pt}
\end{center}

\noindent
$^{1}$Dept.\ of Civil and Environmental Engineering, Carnegie Mellon University, Pittsburgh, PA 15213, USA. Email: \texttt{chenrui2@andrew.cmu.edu}\\
$^{2}$Dept.\ of Civil and Environmental Engineering, Carnegie Mellon University, Pittsburgh, PA 15213, USA. Email: \texttt{bakinci@andrew.cmu.edu}\\
$^{3}$Dept.\ of Mechanical Engineering, Carnegie Mellon University, Pittsburgh, PA 15213, USA. Email: \texttt{ccm@cmu.edu}\\

\section*{ABSTRACT}
AI is increasingly used to support decarbonization decisions across the built environment, yet the development, training, and use of AI consume energy and induce CO$_2$e emissions.
However, existing assessments often report physical-system savings while omitting AI-side emissions. Moreover, they rarely account for the mismatch between when AI costs occur and when decarbonization benefits materialize, which may be substantial for infrastructure-scale projects. 
To address these issues, we present a time-aware assessment framework that models avoided emissions and AI-induced emissions as discrete-time streams over a finite time horizon. In demonstrating this process, we seek to show that time-aware assessment can support temporal decision-making, identify cases in which accounting for time value of carbon can change preferred rankings relative to time-invariant totals, and explore how decisions may vary with slightly different governance priorities. 
Using four representative interventions with intentionally different temporal profiles (multi-project low-carbon concrete design support, AI-assisted construction logistics, agentic HVAC control, and predictive maintenance), we demonstrate how discounting can change preferred rankings relative to time-invariant totals and supports ranking sensitivity analysis, discounted payback screening, and break-even discount-rate analysis.
We also provide decision guidelines that support go/no-go screening, timing decisions, and minimum ``bang-for-your-buck''  thresholds.
Ultimately, this work contributes a lightweight framework for deciding whether and when to deploy AI-enabled interventions for decarbonization under explicit time preference.

\noindent\textbf{Keywords:} artificial intelligence; decarbonization; time value of carbon; environmental net present value; compute footprint; governance

\section*{INTRODUCTION}

AI is increasingly proposed as a means of accelerating decarbonization across the built environment, including both embodied-carbon choices during design and construction and operational-carbon management during use \citep{Rolnick2022TacklingClimateML}.
However, AI is not impact-free: development, training, inference, and the supporting data and compute pipelines consume energy and induce greenhouse gas emissions \citep{Strubell2019EnergyPolicyNLP,Henderson2020ReportingMLFootprints}.
This creates a fundamental tension: an AI deployment may reduce physical-system emissions while adding AI-induced emissions that must be consistently attributed and compared.

Two gaps limit decision support in practice.
First, applied AI-for-decarbonization studies often omit AI-side emissions or report them inconsistently, complicating comparison across interventions with different compute profiles and operational cadence \citep{Ligozat2022HiddenImpactsAI}.
Second, decarbonization decisions are inherently time-dependent: AI costs and benefits often occur on different timelines (e.g., front-loaded training vs.\ delayed savings; ramp-up adoption vs.\ early savings), and emissions factors can change as electricity grids decarbonize \citep{EPA2025eGRIDDetailedData}.
If timing is ignored, options can be misranked and decision making cannot be applied transparently.  Building on prior work that emphasizes the time value of carbon, time preference is therefore treated here as a governance choice rather than a purely technical consideration \citep{Sproul2019TimeValueGHG}.

Related work spans embodied-carbon decision support and uncertainty-aware assessment frameworks \citep{PozzerRauschLeite2024i3CE_EmbodiedCarbonUncertainty,EissaElAdaway2024i3CE_CircularEconomy} as well as AI for building operations (e.g., anomaly detection) \citep{GuWangJazizadeh2024i3CE_TransformersAD}. 
However, these studies often either omit AI-induced emissions or treat emissions outcomes as time-invariant totals, motivating the time-aware accounting and decision metrics presented here.

This paper presents an initial time-aware assessment framework that represents avoided emissions attributable to an AI-enabled intervention, $S(t)$, and AI-induced emissions attributable to development, training, and inference, $C(t)$, as time-indexed streams over a finite horizon. The goal is to enable transparent comparison when costs and benefits occur on different timelines and under explicit time preference.

Using intentionally stylized representative cases with different temporal profiles, we illustrate how discounting can change preferred rankings relative to time-invariant totals and compute break-even discount rates at which rankings reverse. We make three core contributions:
(1) a compact, ISO-aligned time-indexed accounting setup for $S(t)$ and $C(t)$ (\citealp{ISO14040}; \citealp{ISO14044});
(2) time-aware decision metrics (discounted return on carbon, NEPV, discounted carbon payback, and break-even discount rates) with demonstrated usage; and
(3) heuristics for go/no-go screening, timing decisions, and ``bang-for-your-buck'' thresholds.

\section*{METHODS} \label{sec:methods}
This section specifies the details of the assessment process, including the approach for representing avoided and AI-induced emissions as time-indexed streams, and the representative scenarios used for demonstration and exploration.

\subsubsection*{Time-indexed emissions streams and systems boundary}
We consider a finite horizon $t=0,1,\ldots,H$ in discrete annual bins, where $t=0$ denotes one-time deployment activities (e.g., training and integration) and $t=1,\ldots,H$ are subsequent years.
For each intervention $j$, we define a stream of avoided emissions $S_j(t)$ and AI-induced emissions $C_j(t)$ in terms of tCO$_2$e.
$S_j(t)$ can be derived from reduced energy use and time-varying emissions factors, and $C_j(t)$ from training and inference energy and compute-side emissions factors.
The procedure is boundary-aware, including processes that can reasonably change due to AI deployment and excluding components that are identical with or without AI, consistent with ISO 14040/14044 (\citealp{ISO14040}; \citealp{ISO14044}). 
We further represent the time value of carbon using a single environmental discount rate $r$ that is applied consistently across all subsequent calculations. In the illustrative calculations below, we set $r=3\%$ as a reference case consistent with U.S. regulatory analysis guidance \citep{OMB2024A94AppendixC}.

\subsubsection*{Assumptions}
We assume impacts can be represented as additive discrete-time streams $S(t)$ and $C(t)$ over a finite horizon (annual bins here, but monthly or finer bins can be applied as well), and that portfolio impacts are additive (streams can be made summative across interventions). We further assume a constant discount rate $r$ applied uniformly and treat emissions factors as exogenous inputs. All quantities are defined relative to a no-AI baseline and include only processes expected to change due to AI deployment (serving as an attributional, ISO-aligned boundary).

\subsubsection*{Time-aware decision metrics}
The framing given in the previous section permits the calculation of net present value for both emissions avoided through the use of AI, $\mathrm{NPVS}$, and emissions induced through the use of AI, $\mathrm{NPVC}$:

\begin{multicols}{2}
\noindent
\begin{equation}
\label{eq:npvs}
\mathrm{NPVS}(j; r; H)=\sum_{t=0}^{H} \frac{S_j(t)}{(1+r)^t}
\end{equation}
\begin{equation}
\label{eq:npvc}
\mathrm{NPVC}(j; r; H)=\sum_{t=0}^{H} \frac{C_j(t)}{(1+r)^t}
\end{equation}
\end{multicols}

Based on these quantities, we define metrics directly intended to support decision-making. First, we further define a return on carbon value:
\begin{equation}
\label{eq:rr}
R(j; r; H)=\frac{\mathrm{NPVS}(j; r; H)}{\mathrm{NPVC}(j; r; H)}.
\end{equation}
\noindent which is the present-value tons of CO$_2$e avoided per present-value ton of CO$_2$e induced by the AI system over horizon $H$. We interpret this as a discounted carbon-return (efficiency) ratio for the use of AI.
We also define $R_{\mathrm{simple}}(j)=R(j; r=0; H)$, representing the undiscounted ratio of avoided emissions to AI-induced emissions over the horizon. For later tabulation, we define the undiscounted totals
$S_{\mathrm{total}}(j)=\sum_{t=0}^{H} S_j(t),$ and $C_{\mathrm{total}}(j)=C_j(0)+\sum_{t=1}^{H} C_j(t)$.

Next, we define a discounted net benefit, $\mathrm{NEPV}$:
\begin{equation}
\label{eq:nepv}
\mathrm{NEPV}(j;r; H)=\mathrm{NPVS}(j; r; H)-\mathrm{NPVC}(j; r; H)
\end{equation}

Finally, we define the discounted carbon payback time $t_{\mathrm{payback}}$ as the earliest time at which discounted net benefit becomes nonnegative:

\begin{equation}
\label{eq:payback}
t^{(j)}_{\mathrm{payback}}=\min\left\{t:\mathrm{NEPV}(j;r;t)\ge 0\right\}.
\end{equation}

\subsubsection*{Representative scenarios with time-sensitive profiles}
\label{sec:cases}

In order to demonstrate this assessment framework we define four stylized interventions spanning build-phase (pre-occupancy) and use-phase (operations), represented as annual streams over an evaluation horizon $H=10$ years (i.e., a multi-project program or pipeline rather than the duration of a single construction project). The cases are B1: low-carbon concrete design support \citep{Ge2022LowCarbonConcreteCOMPASS}; B2: construction logistics and routing \citep{Karmakar2022RoutePlanningBIMITcon}; U1: agentic HVAC supervisory control \citep{Azuatalam2020RLWholeBuildingHVAC}; and U2: predictive maintenance via fault detection \citep{Singh2022ComprehensiveReviewFDDHVAC}. AI-induced emissions are modeled as a one-time cost at $t=0$ due to training and development plus recurring inference costs for $t\ge 1$ (illustrative, order-of-magnitude).

\begin{figure}[!htbp]
\centering
\includegraphics[width=0.9\linewidth]{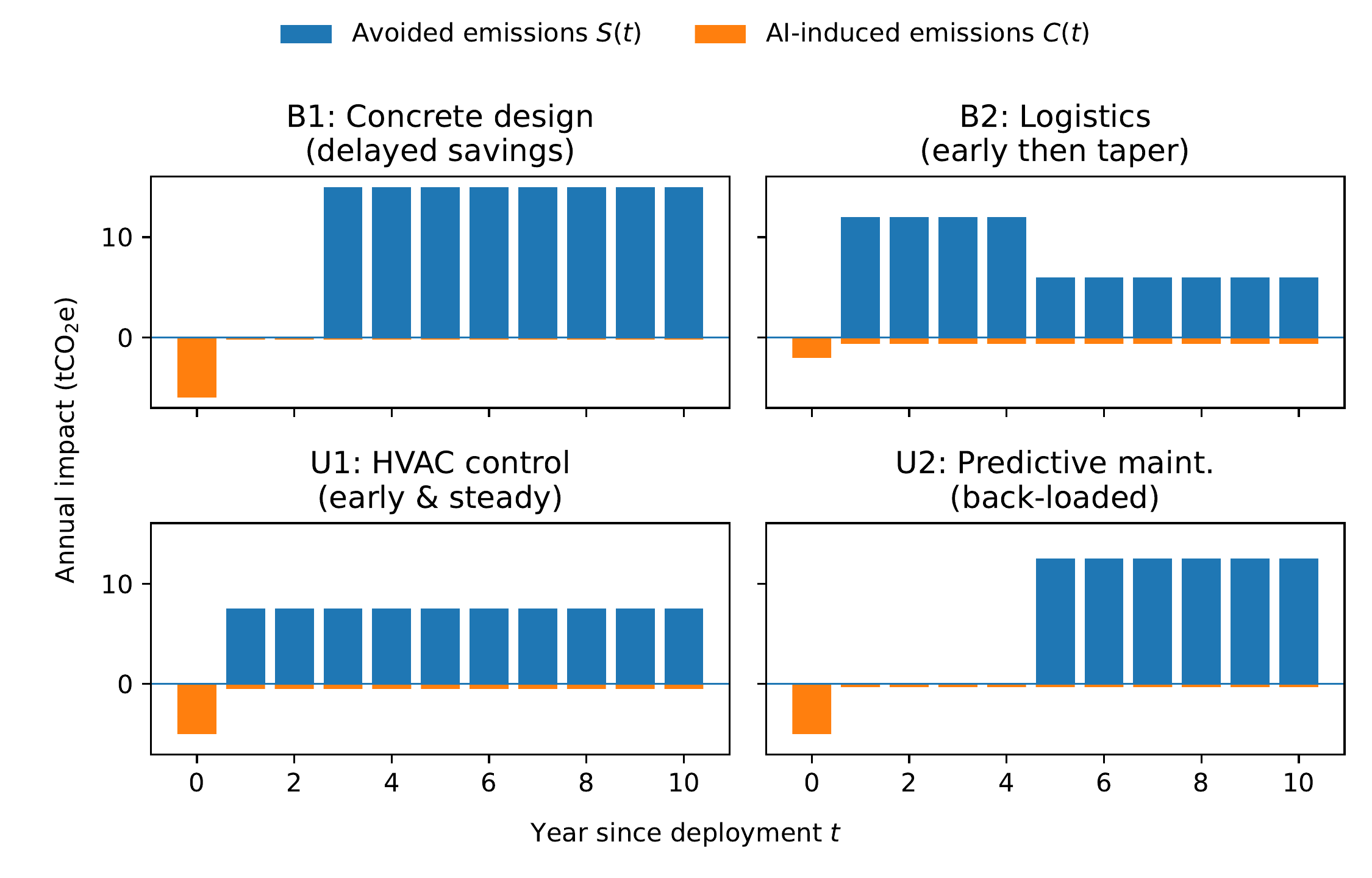}
\caption{Stylized annual streams of avoided emissions $S(t)$ and AI-induced emissions $C(t)$ over horizon $H=10$ (annual bins; $t=0$ is one-time training/integration and $t=1{\dots}H$ are subsequent years). $C(t)$ is plotted as a negative value for readability but treated as a positive value for calculations. For B1 and B2, $S(t)$ is the annual aggregate savings from the new-project cohort in year $t$ under an approximately constant firm pipeline; impacts recur after adoption but do not accumulate across cohorts.}

\label{fig:stream_profiles}
\end{figure}

These streams are parameterized in tCO$_2$e/yr, with one-time training/integration at $t=0$ and recurring inference for $t\ge1$. To address the temporal mismatch between build-phase and use-phase, B1 and B2 are modeled as firm-wide programmatic deployments across a multi-project pipeline rather than to a single building (see Figure~\ref{fig:stream_profiles}).

\begin{itemize}[leftmargin=1em,itemsep=0pt,topsep=1pt]
  \item \textbf{B1 (Concrete pipeline)}: 
  $S(t)=15$ for $t{\ge}3$ and $S(t)=0$ for $1{\le}t{<}3$ (2-yr adoption/integration lag; recurring annual cohort-level savings thereafter);
  $C(0)=6$; $C(t)=0.2$ for $t{\ge}1$.
  
  \item \textbf{B2 (Logistics pipeline)}: 
  $S(t)=12$ for $1{\le}t{\le}4$ (early impact), and $S(t)=6$ for $5{\le}t{\le}10$ (taper);
  $C(0)=2$; $C(t)=0.6$ for $t{\ge}1$.

  \item \textbf{U1 (HVAC control)}: 
  $S(t)=7.5$ for $t{\ge}1$ (early \& steady savings); 
  $C(0)=5$; $C(t)=0.5$ for $t{\ge}1$.

  \item \textbf{U2 (Predictive\ Maintenance)}: 
  $S(t)=12.5$ for $t{\ge}5$ and $S(t)=0$ for $1{\le}t{<}5$ (back-loaded benefit); 
  $C(0)=5$; $C(t)=0.3$ for $t{\ge}1$.
\end{itemize}

\noindent These streams are visualized in Figure~\ref{fig:stream_profiles}. They are intentionally stylized to span distinct timing patterns, with one-time training/integration costs at $t=0$ and recurring inference costs for $t\ge 1$.

\section*{RESULTS}
\label{sec:results}

The results of applying the time-aware metrics to the interventions described above are shown in Table~\ref{tab:single}.
All four cases exhibit time sensitivity because avoided-emissions streams $S(t)$ and AI-induced emissions streams $C(t)$ are temporally mismatched: B1 and U2 are back-loaded (benefits delayed), B2 exhibits tapering benefits over repeated deployments, and U1 combines early/steady savings with persistent inference (see Figure~\ref{fig:stream_profiles}).
Notably, the construction-phase cases (B1 and B2) are sensitive to whether carbon savings are modeled as a multi-year program rather than as a single $t=0$ event. Under time-weighted metrics (e.g., $R$ and payback), this can change rankings even when totals are comparable.

\begin{table}[!htbp]
\centering
\caption{Single-intervention outcomes ($H=10$). Discounted metrics ($R$, NEPV, and $t_{\mathrm{payback}}$) use $r=3\%$; payback is reported in years. Metrics are computed from Eqs.~(\ref{eq:npvs})--(\ref{eq:payback}).}
\label{tab:single}
\small
\begin{tabular}{l >{\raggedright\arraybackslash}p{0.30\linewidth}
                S[table-format=3.0]
                S[table-format=2.1]
                S[table-format=2.2]
                S[table-format=2.2]
                S[table-format=3.2]
                S[table-format=1.0]}
\toprule
Case & Description & {$S_{\mathrm{total}}$} & {$C_{\mathrm{total}}$} & {$R_{\mathrm{simple}}$} & {$R$} & {NEPV} & {$t_{payback}$}\\
\midrule
B1 & Concrete design (delayed savings)            & 120 & 8.0  & 15.00 & 12.88 & 91.54 & 3\\
B2 & Logistics (early impact, then taper)         &  84 & 8.0  & 10.50 & 10.32 & 66.37 & 1\\
U1 & HVAC control (early \& steady savings)       &  75 & 10.0 &  7.50 &  6.91 & 54.71 & 1\\
U2 & Predictive maintenance (back-loaded benefit) &  75 & 8.0  &  9.38 &  7.96 & 52.60 & 5\\
\bottomrule
\end{tabular}

\vspace{1pt}
\raggedright 
\footnotesize 
\textit{Note:} $S_{\mathrm{total}}=\sum_{t=0}^{H} S(t)$ and $C_{\mathrm{total}}=C(0)+\sum_{t=1}^{H} C(t)$ are undiscounted totals from the listed streams.

\normalsize
\end{table}

Table~\ref{tab:portfolio} extends this analysis to compare simple two-intervention portfolios (one build-phase and one use-phase).
The preferred portfolio depends on the governance objective.
In this example, a static ratio favors pairing the high-total, back-loaded options, whereas discounted net impact (NEPV) favors pairings with earlier impacts and shorter payback.

\begin{table}[!htbp]
\centering
\caption{Portfolio outcomes. Discounted metrics ($R$, NEPV, and $t_{\mathrm{payback}}$) use $r=3\%$; payback is reported in years. Portfolio streams are formed by summing constituent $S(t)$ and $C(t)$ streams listed above.}

\label{tab:portfolio}
\small
\begin{tabular}{l >{\raggedright\arraybackslash}p{0.36\linewidth}
                S[table-format=2.2]
                S[table-format=2.2]
                S[table-format=3.2]
                S[table-format=1.0]}
\toprule
Portfolio & Description & {$R_{\mathrm{simple}}$} & {$R$} & {NEPV} & {$t_{payback}$}\\
\midrule
B1 + U1 & Concrete design + HVAC control              & 10.83 &  9.62 & 146.26 & 2\\
B1 + U2 & Concrete design + predictive maintenance    & 12.19 & 10.44 & 144.15 & 3\\
B2 + U1 & Logistics + HVAC control                    &  8.83 &  8.39 & 121.08 & 1\\
B2 + U2 & Logistics + predictive maintenance          &  9.94 &  9.11 & 118.97 & 1\\
\bottomrule
\end{tabular}
\normalsize
\end{table}

The discount rate used in analyses such as these is usually estimated with some degree of error. Therefore, it can be important to examine the critical discount rate at which a decision would reverse.
Here, we compute both the break-even discount rate $r^\star_{R}$ such that $R(j; r^\star_{R}; H)=R(k; r^\star_{R}; H)$, and the break-even discount rate $r^\star_{\mathrm{NEPV}}$ such that $\mathrm{NEPV}(j; r^\star_{\mathrm{NEPV}}; H)=\mathrm{NEPV}(k; r^\star_{\mathrm{NEPV}}; H)$.

For the build-phase comparison (B1 vs.\ B2), the efficiency-based break-even is $r^\star_R=0.0821$. Below this value, the delayed but larger total savings in B1 dominate but above it the earlier-impact B2 becomes more efficient after discounting. Using discounted net impact, the corresponding reversal occurs at $r^\star_{\mathrm{NEPV}}=0.1655$, indicating that B1 remains preferred by $\mathrm{NEPV}$ across a wide range of values. For the use-phase comparison (U1 vs.\ U2), the efficiency-based break-even is similar ($r^\star_R=0.0820$), but the net-impact break-even is much smaller ($r^\star_{\mathrm{NEPV}}=0.0134$~yr$^{-1}$). This means that even modest discounting may favor the earlier-benefit HVAC-control case (U1) over the back-loaded predictive-maintenance case (U2) when decisions are made using $\mathrm{NEPV}$.

\subsubsection*{Worked example for facility management}
This assessment procedure can also be applied in facility management settings to screen AI-enabled
fault detection and supervisory control deployments, where faults persist over time and earlier
fault detection avoids irrecoverable emissions.
Consider an illustrative campus lab-building proxy with annual electricity use
$E_{\mathrm{elec}}=15$~GWh and annual thermal use $E_{\mathrm{therm}}=100{,}000$~MMBtu.
Using a PJM-average, location-based annual emissions factor
$EF_{\mathrm{grid}}=0.40$~kgCO$_2$e/kWh (eGRID) and a thermal proxy factor
$EF_{\mathrm{gas}}=53.06$~kgCO$_2$/MMBtu \citep{EPA2025eGRIDDetailedData},
conservative operational targets of 8\% electricity savings and 5\% thermal savings yield an
avoided-emissions rate
\[
S_{\mathrm{annual}} \approx \frac{0.08\,E_{\mathrm{elec}}\,EF_{\mathrm{grid}}}{1000}
+ \frac{0.05\,E_{\mathrm{therm}}\,EF_{\mathrm{gas}}}{1000}
\approx 745~\mathrm{tCO_2e/yr}.
\]
Assuming continuous cloud inference consumes $E_{\mathrm{comp}}=6{,}000$~kWh/yr, the recurring AI-side
emissions rate is
\[
C_{\mathrm{annual}} \approx \frac{E_{\mathrm{comp}}\,EF_{\mathrm{grid}}}{1000}
\approx 2.4~\mathrm{tCO_2e/yr}.
\]
To align with the payback definition in Eq.~\ref{eq:payback}, let $C_0$ denote a one-time AI-induced carbon cost
incurred up front at deployment (e.g., training and integration) at $t=0$.
Using the steady annual rates above, we apply Eq.~\ref{eq:payback} at annual resolution by setting
$S(0)=0$, $C(0)=C_0$, and $S(1)=S_{\mathrm{annual}}$, $C(1)=C_{\mathrm{annual}}$
(with $r=3\%$ as specified in Methods).
For illustration, if $C_0=100~\mathrm{tCO_2e}$ then Eq.~\ref{eq:payback} yields
$t_{\mathrm{payback}}=1$~year (i.e., within the first year), so the go/no-go gate (Guideline 1) is
likely to be satisfied with substantial margin.
As this is currently applied on an annual basis, any sub-year payback is reported as
$t_{\mathrm{payback}}=1$~year.
If sub-annual payback is decision-relevant, the same definitions can be evaluated on finer discrete
intervals (e.g., monthly) without changing the framework.

\section*{DISCUSSION}
\label{sec:discussion}

A key implication of this work is that construction interventions are not inherently time-invariant with respect to carbon impact (see Table~\ref{tab:single}).
If AI-supported embodied-carbon or logistics decisions are treated as one-off events at $t=0$, time-aware metrics degenerate to static accounting.
However, many organizations deploy such tools programmatically across a pipeline of projects; realized embodied/transport savings are then distributed over multiple years, and AI costs (training, integration, recurring inference) may be front-loaded or persistent.
In this setting, the assumed discount rate can change the ranking of alternatives, and discounted payback provides an actionable go/no-go threshold for deployment.

The metrics above are only decision-relevant insofar as they can be operationalized.
We therefore translate them into three implementable governance heuristics (not hypotheses) that support transparent and checkable go/no-go, timing, and minimum efficiency decisions.

\begin{enumerate}
\item \textbf{Go/no-go Decisions.}
Deploy a solution if $\mathrm{NEPV}(j;r;H)>0$ (projected net carbon savings over horizon $H$ are positive) and $t_{\mathrm{payback}}\le T_{\max}$, where $T_{\max}$ is a governance-specified maximum acceptable payback time (years).

\item \textbf{Timing Decisions.}
If deployment can be delayed by $d\in\{0,1,\ldots,H\}$ years, define shifted streams
$S_{j,d}(t)=S_j(t-d)$ and $C_{j,d}(t)=C_j(t-d)$ with the convention $S_j(\tau)=C_j(\tau)=0$ for $\tau<0$.
Evaluate
\[
\mathrm{NEPV}_d(j;r;H)=\sum_{t=0}^{H}\frac{S_{j,d}(t)-C_{j,d}(t)}{(1+r)^t},
\]
and select $d^\star=\arg\max_d \mathrm{NEPV}_d(j;r;H)$.

\item \textbf{``Bang-for-your-buck'' Decisions.}
Compute the discounted return on carbon $R$.
Adopt a minimum acceptable ratio $R_{\min}$ (set by the decision-maker), and proceed only if $R(j;r;H)\ge R_{\min}$.
This treats $R$ as a simple screening metric for whether the expected avoided emissions are sufficiently large relative to AI-induced emissions at the chosen time preference.

\end{enumerate}

We made two assumptions to simplify the framework: $r$ is constant, and emissions factors are deterministic. For $r$, the break-even analysis above already reports the values at which rankings reverse, so this sensitivity is explicit. The emissions-factor assumption is harder. Real grid emissions vary hour by hour with the generator mix, and annual averages conceal this variation. A fuller treatment would use hour-resolved marginal emissions factors and represent uncertain inputs with probability distributions \citep{Henderson2020ReportingMLFootprints}.

\section*{CONCLUSION}

We presented a time-aware assessment framework for deciding whether and when AI should be deployed for decarbonization.
By representing avoided emissions and AI-induced emissions as time-indexed streams, we demonstrated a series of metrics and computable heuristics for understanding the carbon cost of AI-driven decarbonization efforts.
This type of assessment is critical because AI-side costs are often front-loaded (largely due to training and development) while physical-system benefits often arrive later.

Across four representative interventions, we demonstrated complementary decision metrics and showed that variability in discount-rate choices can reverse rankings.
To translate these metrics into implementable practice, we provided computable heuristics that support (i) transparent go/no-go screening via NEPV and payback thresholds, (ii) explicit timing decisions by evaluating delayed deployment alternatives, and (iii) return on carbon screening through a minimum acceptable threshold. Together, these elements support transparent and checkable deployment decisions and can be applied across AI modalities whenever a minimal set of information (e.g., $S(t)$, $C(t)$, $r$) can be estimated and monitored.

It should be noted that we used stylized scenarios in this work to illustrate the behavior of this assessment framework and the potential sensitivity to discount rate. Therefore, these scenarios should not be treated as a rigorous benchmark evaluation.
Our results inherently depend on attributional assumptions, time-varying emissions factors, and realized savings.
Ongoing work extends this framework to a real-world case study of AI-enabled decarbonization at Carnegie Mellon University's campus facilities to empirically validate framework behavior under observed data.
Future work should also extend this mathematical procedure by incorporating uncertainty and higher-resolution marginal emissions \citep{Henderson2020ReportingMLFootprints}.

\renewcommand{\refname}{REFERENCES}


\begin{thebibliography}{16}

\bibitem[Azuatalam et~al.(2020)]{Azuatalam2020RLWholeBuildingHVAC}
Azuatalam, D., W.-L. Lee, F. de Nijs, and A. Liebman. 2020.
``Reinforcement learning for whole-building HVAC control and demand
response.'' Energy and AI 2: 100020.
\url{https://doi.org/10.1016/j.egyai.2020.100020}.

\bibitem[Eissa and El-Adaway(2024)]{EissaElAdaway2024i3CE_CircularEconomy}
Eissa, R., and I. El-Adaway. 2024. ``Circular economy strategies for
reducing embodied carbon in US commercial building stocks: A system
dynamics modeling approach.'' In Computing in Civil Engineering 2023:
Resilience, Safety, and Sustainability, 729--737. Reston, VA: ASCE.
\url{https://doi.org/10.1061/9780784485248.088}.

\bibitem[EPA(2025)]{EPA2025eGRIDDetailedData}
EPA (U.S. Environmental Protection Agency). 2025. eGRID Detailed Data:
eGRID with 2023 Data. Washington, DC: U.S. EPA.

\bibitem[Ge et~al.(2022)]{Ge2022LowCarbonConcreteCOMPASS}
Ge, X., R. T. Goodwin, H. Yu, P. Romero, O. Abdelrahman, A. Sudhalkar,
J. Kusuma, R. Cialdella, N. Garg, and L. R. Varshney. 2022.
``Accelerated design and deployment of low-carbon concrete for data
centers.'' In Proc. 4th ACM SIGCAS/SIGCHI Conf. on Computing and
Sustainable Societies (COMPASS '22), 340--352. New York: ACM.
\url{https://doi.org/10.1145/3530190.3534817}.

\bibitem[Gu et~al.(2024)]{GuWangJazizadeh2024i3CE_TransformersAD}
Gu, Y., X. Wang, and F. Jazizadeh. 2024. ``Are transformers effective
for time series anomaly detection? A case study in building energy
management.'' In Computing in Civil Engineering 2024: Artificial
Intelligence, Automation and Robotics, and Human-Centered Innovations,
1--11. Reston, VA: ASCE. \url{https://doi.org/10.1061/9780784486115.001}.

\bibitem[Henderson et~al.(2020)]{Henderson2020ReportingMLFootprints}
Henderson, P., J. Hu, J. Romoff, E. Brunskill, D. Jurafsky, and J.
Pineau. 2020. ``Towards the systematic reporting of the energy and
carbon footprints of machine learning.'' J. of Machine Learning Research
21 (248): 1--43.

\bibitem[ISO(2006a)]{ISO14040}
ISO (International Organization for Standardization). 2006a. ISO 14040:
Environmental management---Life cycle assessment---Principles and
framework. Geneva: ISO.

\bibitem[ISO(2006b)]{ISO14044}
ISO (International Organization for Standardization). 2006b. ISO 14044:
Environmental management---Life cycle assessment---Requirements and
guidelines. Geneva: ISO.

\bibitem[Karmakar et~al.(2022)]{Karmakar2022RoutePlanningBIMITcon}
Karmakar, A., A. R. Singh, and V. S. Kumar Delhi. 2022. ``Automated
route planning for construction site utilizing building information
modeling.'' J. of Information Technology in Construction 27: 827--844.
\url{https://doi.org/10.36680/j.itcon.2022.040}.

\bibitem[Ligozat et~al.(2022)]{Ligozat2022HiddenImpactsAI}
Ligozat, A.-L., J. Lefevre, A. Bugeau, and J. Combaz. 2022. ``Unraveling
the hidden environmental impacts of AI solutions for environment.''
Sustainability 14 (9): 5172.

\bibitem[OMB(2024)]{OMB2024A94AppendixC}
OMB (Office of Management and Budget). 2024. OMB Circular No. A-94,
Appendix C: Discount Rates for Cost-Effectiveness, Lease Purchase, and
Related Analyses (Revised November 14, 2024). Washington, DC: White
House.

\bibitem[Pozzer et~al.(2024)]{PozzerRauschLeite2024i3CE_EmbodiedCarbonUncertainty}
Pozzer, A. E., C. Rausch, and F. Leite. 2024. ``Addressing uncertainty
in embodied carbon assessment through stochastic analysis to support
decision-making in construction projects.'' In Computing in Civil
Engineering 2024: Sustainability, Resilience, Safety, and Education.
Reston, VA: ASCE. \url{https://doi.org/10.1061/9780784486139.037}.

\bibitem[Rolnick et~al.(2022)]{Rolnick2022TacklingClimateML}
Rolnick, D., P. L. Donti, et al. 2022. ``Tackling climate change with
machine learning.'' ACM Computing Surveys.

\bibitem[Singh et~al.(2022)]{Singh2022ComprehensiveReviewFDDHVAC}
Singh, V., J. Mathur, and A. Bhatia. 2022. ``A comprehensive review:
Fault detection, diagnostics, prognostics, and fault modeling in HVAC
systems.'' International Journal of Refrigeration 140: 1--24.
\url{https://doi.org/10.1016/j.ijrefrig.2022.05.021}.

\bibitem[Sproul et~al.(2019)]{Sproul2019TimeValueGHG}
Sproul, E., J. Barlow, and J. C. Quinn. 2019. ``Time value of greenhouse
gas emissions in life cycle assessment and techno-economic analysis.''
Environmental Science \& Technology 53 (10): 6073--6080.
\url{https://doi.org/10.1021/acs.est.9b00514}.

\bibitem[Strubell et~al.(2019)]{Strubell2019EnergyPolicyNLP}
Strubell, E., A. Ganesh, and A. McCallum. 2019. ``Energy and policy
considerations for deep learning in NLP.'' In Proc. 57th Annual Meeting
of the Association for Computational Linguistics (ACL), 3645--3650.

\end{thebibliography}
\end{document}